\documentstyle[10pt,psfig]{article}
\begin{document}
\begin{center} {\bf Speckle interferometric observations of the collision of
comet Shoemaker-Levy 9 with Jupiter} 
\end{center} 
\vspace{0.7cm}

\noindent
S K Saha$^{1}$, R Rajamohon$^{1}$, P Vivekananda Rao$^{2}$, G Som Sunder$^{2}$, 
R Swaminathan$^{2}$ and B Lokanadham$^{2}$ \\
\vspace{0.5cm}

\noindent
$^{1}$Indian Institute of Astrophysics, Bangalore 560034, India \\
$^{2}$Astronomy Department, Osmania University, Hyderabad 500007, India \\
\vspace{0.5cm}

\noindent
{\bf Abstract} Speckle interferometric technique has been used to obtain a
series of short exposure images of the collision of comet Shoemaker-Levy 9
with Jupiter during the period of July 17-24, 1994 using the Nasmyth focus
of 1.2 meter telescope of Japal-Rangapur Observatory, Hyderabad. The technique 
of Blind Iterative Deconvolution (BID) was used to remove the atmospherically 
induced point spread function (PSF) from these images to obtain diffraction 
limited informations of the impact sites on Jupiter.
\vspace{0.4cm}

\noindent
Key words: Speckle Imaging, Image Reconstruction, Jupiter, Shoemaker-Levy 9
\vspace{0.4cm}

\begin{center} 
{\bf 1. Introduction} 
\end{center} 
\vspace{0.4cm}

\noindent
The impact of the collision of the comet Shoemaker-Levy 9 (1993e) with the 
gaseous planet Jupiter during the period 16th.-22nd. July, 1994, has been
observed extensively worldwide, as well as from the Hubble space Telescope.
Several observatories in India too had planned observations of the crash
phenomena starting from the observations in the visible part of the 
electro magnetic spectrum to the radio frequencies (Cowsik, 1994). As a part 
of the programmes, we had developed an interferometer to record the images
of the collision of the fragments of Comet Shoemaker-Levy 9 (SL 9) with 
Jupiter during the period 17-24th. July, 1994, with a goal of achieving 
features with a resolution of 0.3-0.5 arc sec., in the optical band, 
using 1.2 meter telescope at Japal-Rangapur Observatory (JRO), Osmania 
University, Hyderabad. Though, monsoon condition prevailed over large part of 
the country, we were able to record more than 600 images of the entire 
planetary disk of Jupiter during the said period. In this paper, we describe 
the observational technique using interferometer, as well as the image 
processing technique used to restore the degraded images of Jupiter.
\vspace{0.4cm}

\begin{center} 
{\bf 2. Observations} 
\end{center}
\vspace{0.4cm}

\noindent
The image scale at the Nasmyth focus (f/13.7) of 1.2 meter telescope
of JRO, was enlarged by a Barlow lens arrangement (Saha {\it et al.}, 1987,
Chinnappan {\it et al.}, 1991). The set up was modified to suit to requirement
of sampling 0.11 arc sec/pixel of the CCD (at 0.55 $\mu$) which
is essentially the diffraction limit of the said telescope. A set of
3 filters were used to image Jupiter, viz.,:
(i) centered at 5500 $\AA$, with FWHM of 300 $\AA$,
(ii) centered at 6110 $\AA$, with FWHM of 99 $\AA$, and
(iii) RG9 with a lower wavelength cut-off at 8000 $\AA$.
\vspace{0.2cm}

\noindent
A 1024$\times$1024 pixel water cooled CCD with a pixel size 22$\mu$ was used
as a detector. 50 speckle-grams were sequentially recorded (each of 100 m sec
exposure) in each of the 3 filters. The exposure time was chosen to obtain a
good signal-to noise ratio. Since the smearing due to the equatorial rotation
of Jupiter is about 0.15 arc sec/min., one can afford to accumulate speckle-grams
for 2-3 minutes, if one expected to attain a resolution of 0.5 arc sec.
In this experiment, we have recorded 10 speckle-grams/min. Therefore,
20-30 frames with good enough signal-to-noise ratio at the desired spatial
frequencies are required to perform speckle reconstruction. 600 images were
recorded on July 17, 1994 soon after the fragment E of the Comet SL-9 collided
with Jupiter. On July 24, 1994, 80 more images were recorded. A liquid
nitrogen cooled 512$\times$512 CCD was used to record 3 images of Jupiter in 
integrated light on July 22, 1994.
\vspace{0.4cm}
 
\begin{center} 
{\bf 3. Data Processing} 
\end{center}
\vspace{0.3cm}

\noindent
Atmospherically induced phase fluctuations distort incoming plane wave-fronts
from the distant objects which reach the entrance pupil of telescope with 
patches of random excursions in phase. Such phase distortions restrict the
effective angular resolution of most telescopes to 1 second of arc or worse.
Speckle interferometry (Labeyrie, 1970) recovers the 
diffraction limited spatial Fourier spectrum and image features of the object
intensity distribution from a series of short-exposure ($<$ 20 m sec.) images. 
Schemes like Knox-Thompson algorithm (Knox-Thompson, 1974),
triple correlation (Lohmann {\it et al.}, 1983) have been successfully
employed to restore the Fourier phase of an extended object. All these schemes 
require statistical treatment of a large number of images. Often, it may not be
possible to record a large number of images within the time interval over which 
the statistics of the atmospheric turbulence remains stationary. 
There are a number of schemes, viz., Maximum Entropy Method (Jaynes, 1982),
CLEAN algorithm (Hogbom, 1974) and Blind Iterative Deconvolution (BID) 
technique (Ayers and Dainty, 1988) being applied to restore the image using 
some prior information about the image. Here, we employed a version of BID 
developed by P. Nisenson (Nisenson, 1991), on degraded images of Jupiter. 
\vspace{0.2cm}

\noindent
In this technique (see Bates and McDonnell, 1986), the iterative
loop is repeated enforcing image-domain and Fourier-domain constraints until
two images are found that produce the input image when convolved together.
The image-domain constraint of non-negativity is generally used in 
iterative algorithms associated with optical processing to find effective
supports of the object and or point spread function (PSF) from a speckle-gram.
Here, the Weiner filter was used to estimate one function from an initial 
guess of the PSF.
\vspace{0.2cm}

\noindent
The algorithm has the degraded image ${c(x,y)}$ as the operand.  An initial
estimate of the point spread function (PSF) ${p(x,y)}$ has to be provided.
The degraded image is deconvolved from the guess PSF by Wiener filtering,
which is an operation of multiplying a suitable Wiener filter (constructed
from the Fourier transform ${P(u,v)}$ of the PSF) with the Fourier transform 
${C(u,v)}$ of the degraded image as follows
\vspace{0.2cm}
 
\begin{math}
{O(u,v) = C(u,v) {\frac {P^{*}(u,v)} 
{P(u,v)P^{*}(u,v) + N(u,v)N^{*}(u,v)}}} 
\end{math} 
\vspace{0.2cm}

where $O$ is the Fourier transform of the Deconvolved image and $N$ is the
noise spectrum. 
\vspace{0.2cm}

\noindent
This result $O$ is transformed to image space, the negatives in the image
are set to zero, and the positives outside a prescribed domain (called
object support) are set to zero. The average of negative intensities
within the support are subtracted from all pixels. The process is repeated
until the negative intensities decrease below the noise.
\vspace{0.2cm}

\noindent 
A new estimate of the PSF is next obtained by Wiener filtering the original
image ${c(x,y)}$ with a filter constructed from the constrained object
${o(x,y)}$. This completes one iteration. This entire process is repeated until 
the derived values of ${o(x,y)}$ and ${p(x,y)}$ converge to sensible solutions.
\vspace{0.4cm}

\begin{center} 
{\bf 4. Results} 
\end{center}
\vspace{0.4cm}

\noindent
The flat field corrections, as well as bias subtractions were made for all
the Jupiter images acquired on 17th. and 24th. July '94, using IRAF image
processing package and analyzed on SPARC ultra Workstation. The images were
converted to specific formats to make them IRAF compatible. The results for the 
Jupiter images obtained on 17th. July '94 and on 24th. July '94 were arrived at 
after 150 iterations. Since the combined PSF of the atmosphere and the telescope 
varies as a function of time, the value of support radius of the PSF had been 
chosen accordingly. The value of the Weiner filter parameters also had chosen 
according to the intensity of each of the images differently. 
Figure 1 shows the speckle-gram of the Jupiter obtained on 24th. July '94,
through the green filter centered at 5500 $\AA$, with FWHM of 300 $\AA$. The 
satellite Io can be seen on top left. Care has been taken to avoid the
satellite while reconstructing the images.
Figure 2 shows the deconvolved image of the same. The complex structure
of the spots were identified and compared with Hubble space telescope
observation. The chief result of this reconstruction is the enhancement in
the contrast of spots. The complex spot at the East is due to impacts by
fragments ${Q_2}$, R, S, D and G. The spot close to the centre is due to
K and L impacts. Figure 3 depicts the reconstructed PSF. 
\vspace{0.2cm}

\noindent
\begin{figure}
\centerline{\psfig{figure=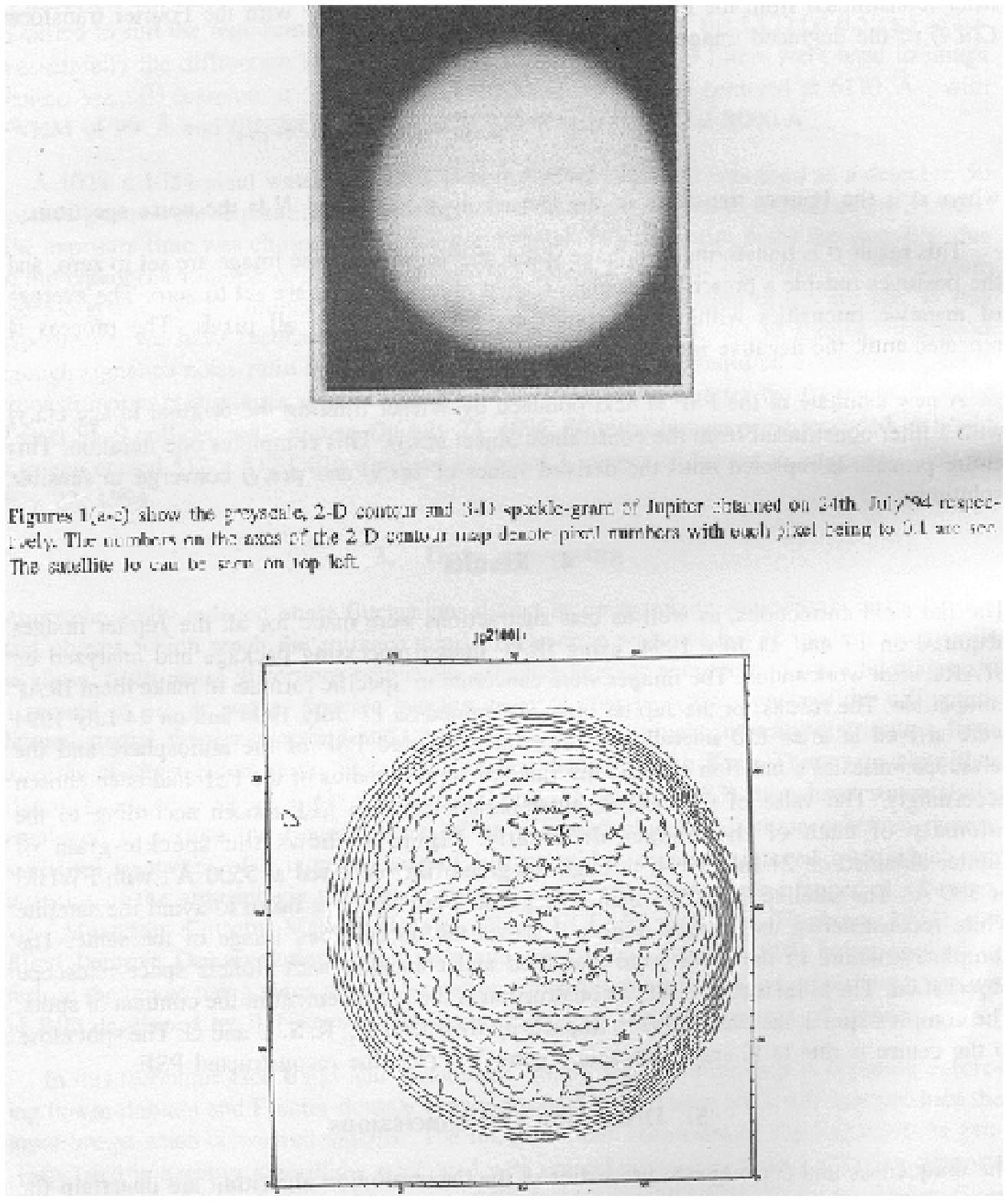,height=14cm,width=12cm}}
\end{figure}

\noindent
\begin{figure}
\centerline{\psfig{figure=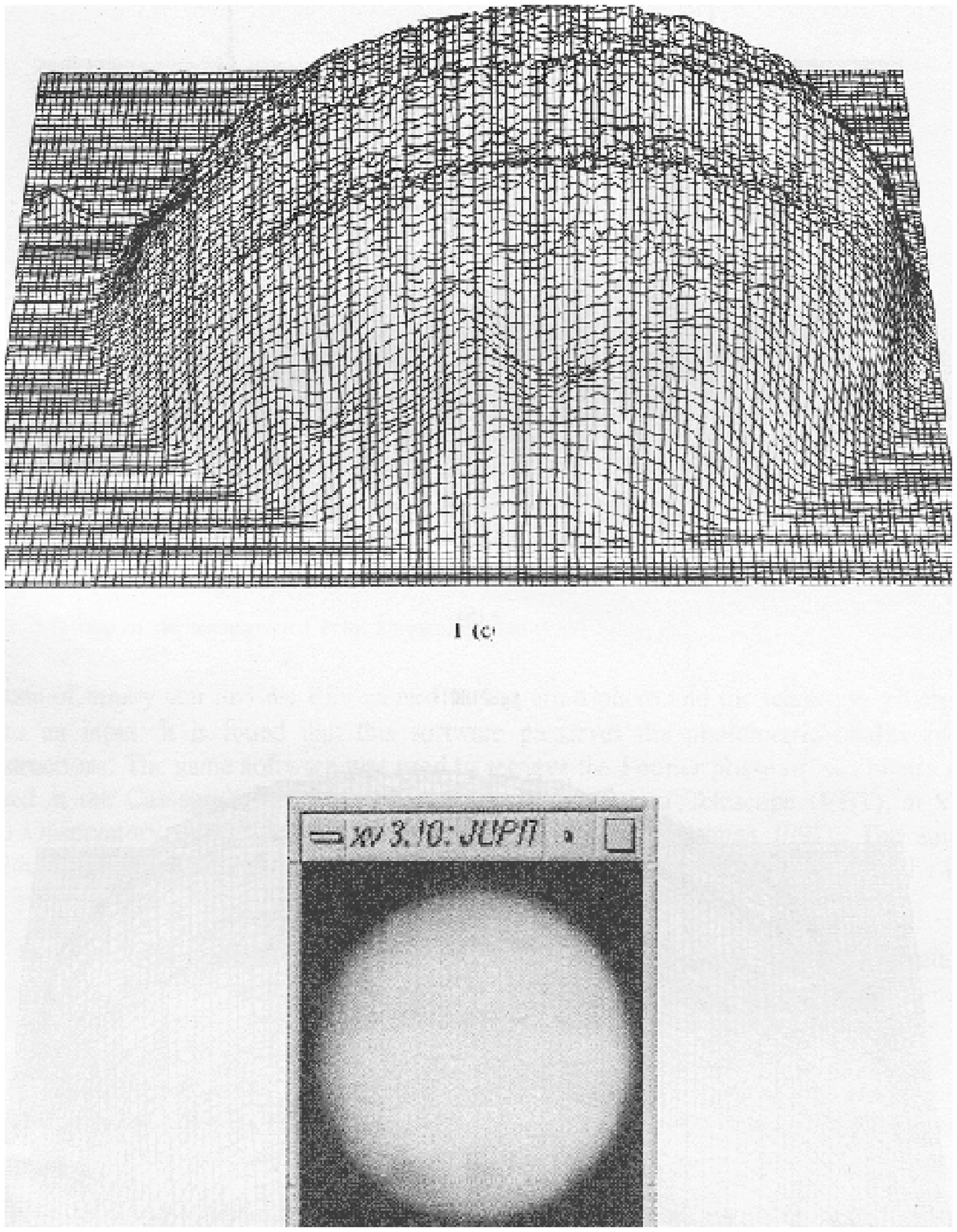,height=14.5cm,width=12cm}}
\end{figure}

\noindent
\begin{figure}
\centerline{\psfig{figure=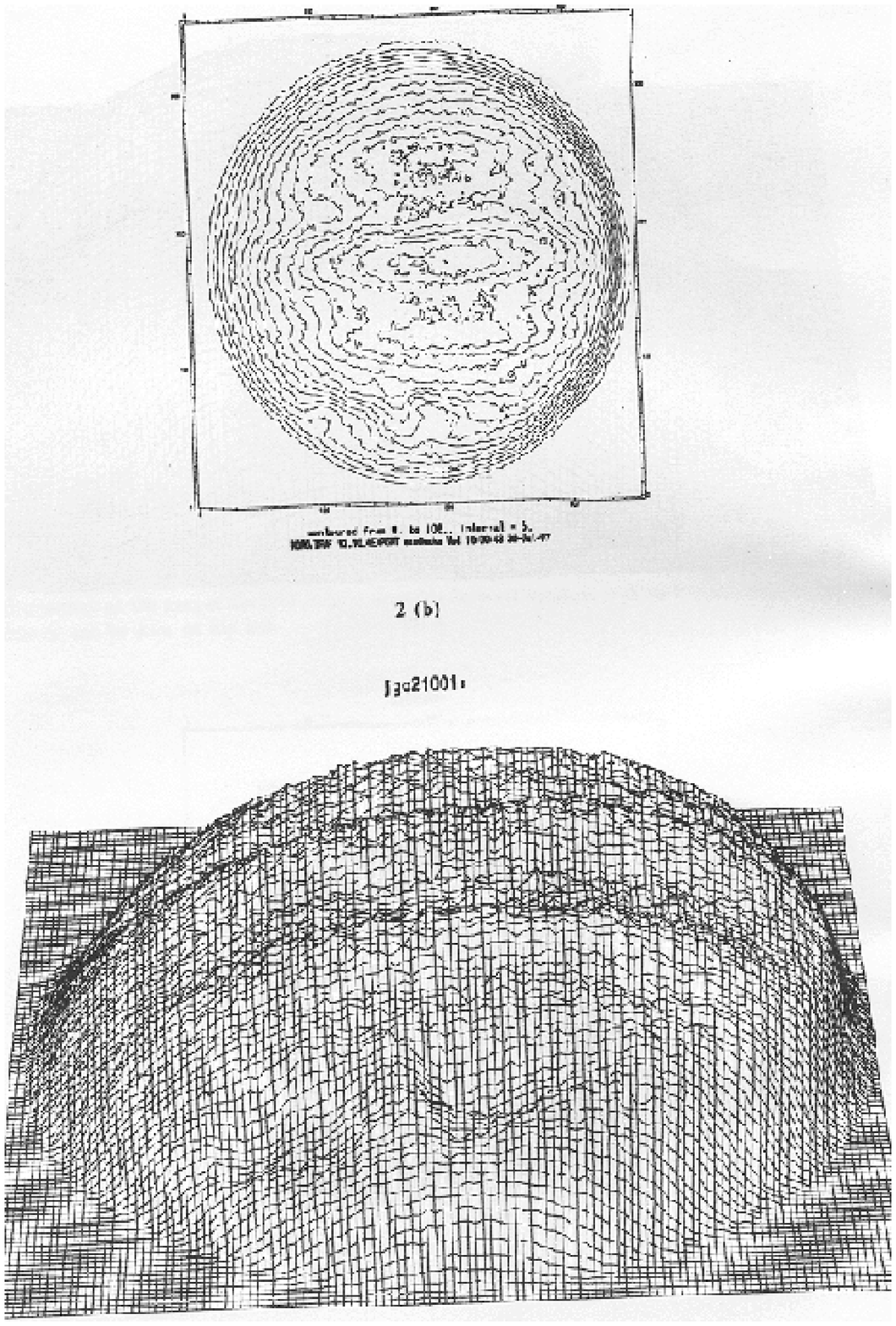,height=17cm,width=12cm}}
\end{figure}

\noindent
\begin{figure}
\centerline{\psfig{figure=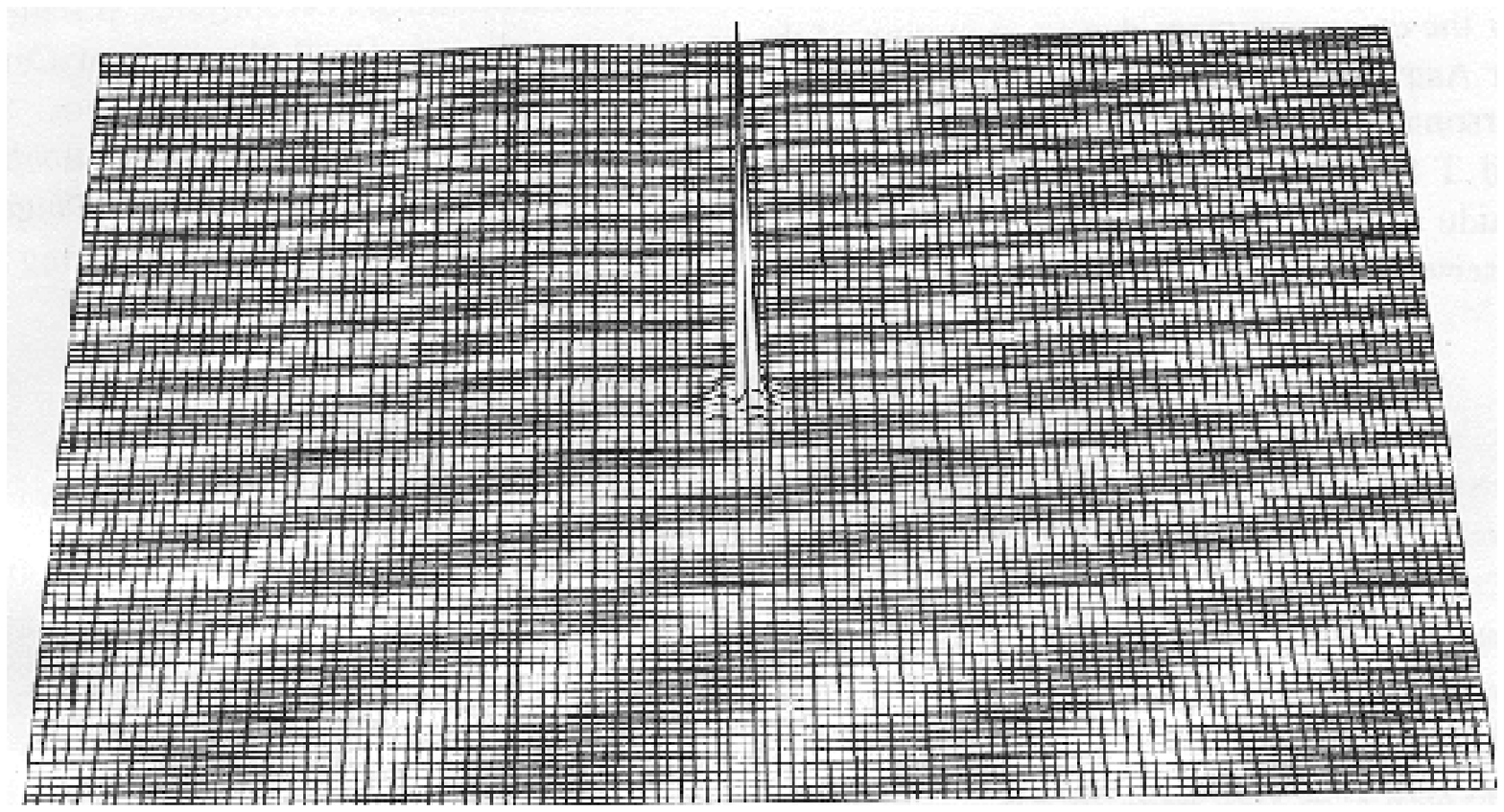,height=7cm,width=12cm}}
\end{figure}

\begin{center}
{\bf 5. Discussion and Conclusions} 
\end{center} 
\vspace{0.4cm}

\noindent
The uniqueness and convergence properties of the Deconvolution algorithm
are uncertain for the evaluation of the reconstructed images if one uses 
BID directly. The support radius of the PSF was estimated from the observations
around the same time. The present scheme of BID has been tested by 
reconstructing the Fourier phase of the computer simulated convolved
functions of binary star and the PSF caused by the atmosphere and the 
telescope which was used as an input. It is found that this software preserves
the photometric quality of the reconstructions. The same software was used to 
retrieve the Fourier phase of two binary stars obtained at the Cassegrain end of 
the 2.34 meter Vainu Bappu Telescope (VBT), at Vainu Bappu Observatory (VBO),
Kavalur, India (Saha and Venkatakrishnan 1997). The authors have found the
magnitude difference for the reconstructed objects compatible with the published
values. 
\vspace{0.2cm}

\noindent
In the case of compact sources, it is essential and possible to put a
support constraint on the object obtained from the auto-correlation. Whereas,
in this case of an extended object of a complex structure, the auto-correlation
is of not great help to constrain the individual features. Convergence could
not be obtained in the absence of any constraints. The estimated support radius
of the PSF was utilized in this case and a successful convergence was 
obtained. Thus in the case of complex objects, the prior knowledge of the
PSF support radius seems to be vital for the reconstructions. 
\vspace{0.2cm}

\noindent
Though the chief problem of this software is that of convergence, it is
indeed an art to decide when to stop the iterations. The results are
also vulnerable to the choice of various parameters like the support
radius, the level of high frequency suppression during the Wiener filtering, 
etc. The availability of prior knowledge on the PSF, in this case, of the 
degraded image was also found to be very useful.  It is to be seen how the 
convergence could be improved (cf.\ Jefferies and Christou, 1993). 
For the present, it is noteworthy that such reconstructions are
possible using single speckle frames. 
\vspace{0.4cm}

\noindent
{\bf Acknowledgments:} The authors are grateful to Prof. R Cowsik, 
Director, Indian Institute of Astrophysics, Bangalore for the encouragement
during execution of the project, as well as to Dr. P. Nisenson
of Center for Astrophysics, Cambridge, USA, for the BID code as well as
for useful discussions. The personnel of the mechanical division of IIA,
in particular Messrs F Gabriel, K Sagayanathan and T Simon, 
helped in the fabrication of the instrument. The help rendered by
B Nagaraj Naidu of electronic division of IIA ,Bangalore and M J Rosario of VBO,
Kavalur, during the observations are gratefully acknowledged.
\vspace{0.4cm}

\begin{center} 
{\bf References} 
\end{center} 
\vspace{0.4cm}

\noindent
Ayers G.R. \& Dainty J.C., 1988, Optics Letters, {\bf 13}, 547. \\
\noindent
Bates R.H.T. \& McDonnell M.J., 1986, "Image Restoration and Reconstruction",
Oxford Engineering Science {\bf 16}, Clarendon Press, Oxford. \\
\noindent
Chinnappan V., Saha S.K. \& Fassehana, 1991, Kod. Obs. Bull. {\bf 11}, 87. \\
\noindent
Cowsik R., 1994, Current Sci., {\bf 67 },      . \\
\noindent
Hogbom J., 1974, Ap.J. Suppl., {\bf 15}, 417. \\
\noindent
Jaynes E.T., 1982, Proc. IEEE, {\bf 70}, 939. \\
\noindent
Jefferies S.M. \& Christou J.C., 1993, ApJ., {\bf 415}, 862. \\
\noindent
Knox K.T. \& Thompson B.J., 1974, Ap.J. Lett., {\bf 193}, L45. \\ 
\noindent
Labeyrie A., 1970, A and A, {\bf 6}, 85. \\
\noindent
Lohmann A.W., Weigelt G. \& Wirnitzer B., 1983, Appl. Opt., {\bf 22}, 4028. \\ 
\noindent
Nisenson P., 1991, in Proc. ESO-NOAO conf. on High Resolution Imaging
by Interferometry ed. J M Beckers \& F Merkle, p-299. \\
\noindent
Saha S.K. \& Venkatakrishnan P., 1997, Bull. Astron. of India (To appear). \\
\noindent
Saha S.K., Venkatakrishnan P., Jayarajan A.P. \& Jayavel N., 1987, Current 
Science, {\bf 56}, 985. \\

\begin{center}
{\bf Figure captions}
\end{center}
\vspace{0.3cm}

\noindent
1. Fig. 1: 1a, 1b, 1c show the greyscale, 2-D contour and 3-D speckle-gram of 
Jupiter obtained on 24th. July '94 respectively. The numbers on the axes
of the 2-D contour map denote pixel numbers with each pixel being equal to
0.1 arc sec. The satellite Io can be seen on top left.
\vspace{0.2cm}

\noindent 
2. Fig. 2: 2a, 2b, 2c are for the Deconvolved image of the Jupiter on 24th. 
July '94.
\vspace{0.2cm}

\noindent
3. Fig. 3: 3-D map of the reconstructed Point Spread Function (PSF). 
 
\end{document}